\documentclass[twocolumn,showpacs,preprintnumbers,amsmath,amssymb]{revtex4}

\usepackage{graphicx}
\usepackage{dcolumn}
\usepackage{bm}

\begin{document}
\title{Emergence of cooperation induced by preferential learning}
\author{Jie Ren$^{1}$}
\author{Wen-Xu Wang$^{2}$}
\author{Gang Yan$^{3}$}
\author{Bing-Hong Wang$^{2}$}
\email{bhwang@ustc.edu.cn}
\affiliation{$^{1}$Department of Physics,
\\$^{2}$Department of Modern Physics,
\\$^{3}$Department of Electronic Science and Technology, \\University of
Science and Technology of China, Hefei, 230026, PR China }

\date{\today}

\begin{abstract}
The evolutionary Prisoner's Dilemma Game (PDG) and the Snowdrift
Game (SG) with preferential learning mechanism are studied in the
Barab\'asi-Albert network. Simulation results demonstrate that the
preferential learning of individuals remarkably promotes the
cooperative behavior for both two games over a wide range of
payoffs. To understand the effect of preferential learning on the
evolution of the systems, we investigate the time series of the
cooperator density for different preferential strength and
payoffs. It is found that in some specific cases two games both
show the $1/f$-scaling behaviors, which indicate the existence of
long range correlation. We also figure out that when the large
degree nodes have high probability to be selected, the PDG
displays a punctuated equilibrium-type behavior. On the contrary,
the SG exhibits a sudden increase feature. These temporary
instable behaviors are ascribed to the strategy shift of the large
degree nodes.

\end{abstract}


\maketitle
\section{Introduction}

Cooperation is ubiquitous in real world, ranging from biological
systems to economic and social systems \cite{cooperation}.
However, the unselfish, altruistic actions apparently contradict
Darwinian selection. Thus, understanding the conditions for the
emergence and maintenance of cooperative behavior among selfish
individuals is a central problem \cite{gene}. Game theory together
with its extensions \cite{von,nash,smith,axelrod,sigmund,http},
considered to be an important approach, provides a useful
framework for investigating this problem. Two simple games,
Prisoners' Dilemma Game (PDG) \cite{PD} and Snowdrift Game (SG)
\cite{SG}, as metaphors for characterizing the evolution of
cooperative behavior have drawn much attention from not only
social but also biological and physical scientists
\cite{bio1,bio2,Nowak1,Nowak2,Nowak3,Kim,Szabo1,Szabo2,Szabo3,Szabo4,Szabo5,Szabo6,Hauert,Doebeli,prl}.
In the original PDG, each of two players may chose either to
cooperate or defect in any one encounter. If they both cooperate,
both obtain a payoff of $R$, whereas mutual defection results in
pay-off $P$ for both players. If one player defects while the
other cooperates, defector gains the biggest pay-off $T$, while
cooperator gets $S$. The ranking of four pay-off values is
$T>R>P>S$. The SG is a game of much biologically interesting. This
game differs from the PDG mainly in the order of $P$ and $S$, as
$T>R>S>P$, which are more favorable to sustain cooperative
behavior.

However, in these two games, the unstable cooperative behavior is
opposite to the observations in the real world. This disagreement
thus motivates a variety of suitable extensions of basic model rules
to explain the emergence of cooperation. Some previous works have
suggested that the introduction of ``tit-for-tat"
\cite{axelrod,Nowak2} strategy can remarkably enhance the
cooperative behavior. More recently, Nowak and May \cite{Nowak1}
found that the PDG with simple spatial structure can induce the
emergence of cooperation, and in particular, spatial chaos is
observed. In contrast, the work of Hauert and Doebeli \cite{Hauert}
demonstrates the spatial structure often inhibits the evolution of
cooperation in the SG. Inspired by the idea of spatial game, much
attention has been given to the interplay between evolutionary
cooperative behavior and the underlying structure
\cite{Szabo1,Szabo2,Szabo3,Szabo6,Kim,prl}. Since the surprising
discovery of ``small world" \cite{SW} and ``scale-free" \cite{BA}
structural properties in real networked systems, evolutionary games
are naturally considered on the networks with these two kinds of
structural features \cite{Szabo4,Szabo5,Szabo6,Kim,prl}.
Interestingly, it is found that comparing with the square lattices,
Scale-free networks provide a unifying framework for the emergency
of cooperation \cite{prl}.

In the two games with network structure, such as square lattices
(spatial structure), small world and scale-free structure, players
interact only with their immediate neighbors. In each round, the
score of each individual is the sum of the payoffs in the
encounters with its neighbors. At the next generation, all the
individuals could update their strategies (cooperate or defect)
synchronously according to either the deterministic rule
introduced by Nowak and May \cite{Nowak1} or the stochastic
evolutionary rule by Szab\'{o} and T\H{o}ke \cite{Szabo1}.

In this paper, we focus on the PDG and SG on scale-free networks
mainly according to the stochastic update rules. However, we argue
that such as in the social system, individual may not completely
randomly choose a neighbor to learn from it. ``Rich gets richer"
is a common feature in social and natural system, which reveals
the existence of preferential mechanism. It is indeed the
preferential attachment mechanism of Barab\'asi and Albert model
(BA for short) \cite{BA} leads to the scale-free structural
property in good accord with the empirical observations. Thus, in
the present work, we present a preferential learning rule, the
probability of  which is governed by a single parameter, for
better mimicking the evolution of real world system.  The
probability of choosing a neighbor for each individual depends on
the degree of that neighbor. This assumption takes into account
that the status of individuals can be reflected by the degree of
them in various communities in nature and society, e.g. the leader
usually interacts with large quantities of individuals.
Interestingly, we find that the preferential learning mechanism
promotes the cooperative behavior of both the PDG and SG. Several
attractive properties for some specific parameter values are
observed, such as the $1/f$-like noise of evolutionary cooperator
density for both two games, which indicates the long range
correlation of cooperation. In the SG, for some specific cases,
the degree of cooperation displays a punctuated equilibrium-type
behavior instead of steady state. In contrast, the PDG exhibits an
absolutely different property of sudden jumps of cooperation.
These two distinct behaviors are both attributed to the effect of
leaders, i.e. the individuals with large connectivity.

The paper is arranged as follows. In the following section we
describe the model in detail, in Sec. III simulations and analysis
are provided for both the PDG and SG , and in Sec. IV the work is
concluded.

\section{The model}
We first construct the scale-free networks using the BA model
which is considered to be the most simple and general one.
Starting from $m_0$ fully connected nodes, one node with $m$ links
is attached at each time step in such a way that the probability
$\Pi_i$ of being connected to the existing node $i$ is
proportional to the degree $k_i$ of that node, i.e.,
$\Pi_i=k_i/\sum_j k_j$ with summation over all the existing nodes.
Here, we set $m=m_0=2$ and network size $N=5000$ for all
simulations. The degree distribution of BA networks follows a
power law $P(k)\sim k^{-3}$\cite{BA}.

We consider the evolutionary PDG and SG on the networks. Without
losing generality, we investigate the simplified games with a
single payoff parameter following previous works
\cite{Nowak1,Szabo1,Hauert}. Figure (1) illustrates the encounter
payoffs of both the PDG and SG. Each individual is placed on a
node of the network and plays the games only with their immediate
neighbors simultaneously. The total payoff of each player is the
sum over all its encounters.

\begin{figure}
\scalebox{0.95}[0.95]{\includegraphics{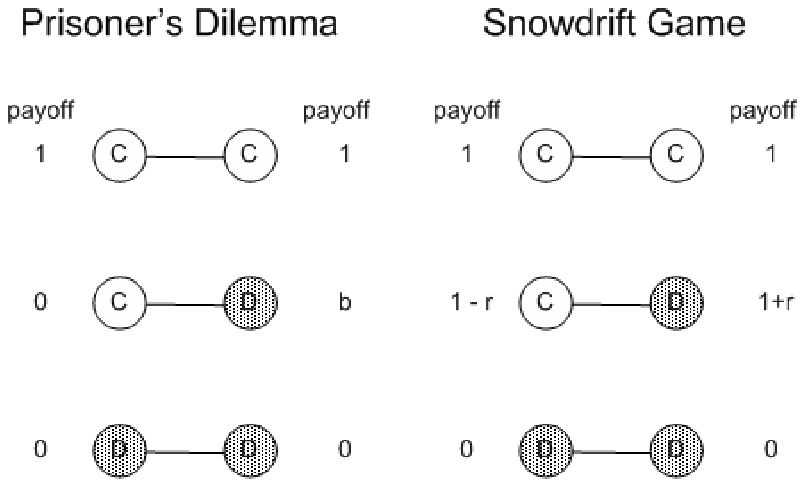}}
\caption{\label{fig:epsart} The payoffs: In the PDG, when two
cooperators $(C)$ encounter, both earn 1. While two defectors
$(D)$ encounter, both earn 0. When a cooperator encounters a
defector, the defector earns $b$ and the cooperator 0. In the SG,
it is the same as PDG when two cooperators or defectors encounter.
However, when a cooperator meets a defector the cooperator scores
$1-r$ and the defector scores $1+r$.}
\end{figure}

During the evolutionary process, each player is allowed to learn
from one of its neighbors and update its strategy in each round. As
mentioned early, each player chooses a neighbor according to the
preferential learning rule, i.e., the probability $P_{i\rightarrow
j}$ of $i$ selecting a neighbor $j$ is
\begin{equation}
P_{i\rightarrow j}=\frac{k_i^\alpha}{\sum_jk_j^\alpha},
\end{equation}
where $\alpha$ is a tunable parameter and the sum runs over the
neighbors of $i$. One can see when $\alpha$ equals zero, the
neighbor is randomly selected so that the game is reduced to the
original one. While in the case of $\alpha
>0$, the individuals with large degree have advantages to be
selected; Otherwise, the small degree individuals have larger
probability to be selected. In social and natural systems, some
individuals with high status and reputation may have much stronger
influence than others and the status of individuals can be reflected
by the degree of them. Thus, the introduction of the preferential
learning intends to characterize the effect of influential
individuals on the evolution of cooperation. In parallel, we also
investigate the performance of the systems with tendency of learning
from the individuals with small degree.

After choosing a neighbor $y$, the player $x$ will adopt the
coplayer's strategy with a probability depending on the normalized
total payoff difference presented in Ref. \cite{Szabo5} as
\begin{equation}
W=\frac{1}{1+\exp[(M_x/k_x-M_y/k_y)/T]},
\end{equation}
where $M_x$ and $M_y$ are the total incomes of player $x$ and $y$,
and $T$ characterizes the noise effects,including fluctuations in
payoffs, errors in decision, individual trials, etc. This choice
of $W$ takes into account the fact of bounded rationality of
individuals in sociology and reflects natural selection based on
relative fitness in terms of evolutionism. The ratio of total
income of individual and its degree, i.e., $M_x/k_x$ denotes the
normalized total payoff. This normalization avoids an additional
bias from the different degree of nodes. In the next section, we
perform the simulations of the PDG and SG respectively and our
goal is to find how the preferential learning affect the
evolutionary cooperative behaviors of both PDG and SG.

\begin{figure}
\scalebox{0.80}[0.80]{\includegraphics{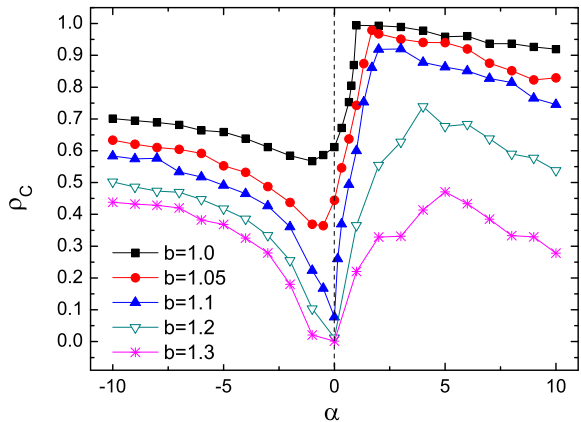}}
\caption{\label{fig:epsart} (color online). The cooperator density
$rho_C$ of the PDG as a function of parameter $\alpha$ for
different value of $b$. }
\end{figure}

\section{simulation results}
The key quantity for characterizing the cooperative behavior of
the system is the density of cooperators $\rho_C$. Hence, we first
investigate $\rho_c$ as a function of the tunable parameter
$\alpha$ for different payoff parameter $b$ in the PDG, as shown
in Fig. 2. The simulation results were obtained by averaging over
last $10000$ time steps of entire $20,000$ time steps. Each data
point results from an average over $20$ simulations for the same
type of network structure. In the initial state, the strategies of
$C$ and $D$ are uniformly distributed among all the players. We
figure out that, comparing with the case of no preferential
learning, i.e.,$\alpha=0$, the cooperation is remarkably promoted
not only for positive value of $\alpha$, but also for negative
$\alpha$ in a wide range of $b$. For negative $\alpha$, the
$\rho_c$ monotonously increases with the decrease of $\alpha$ and
finally $\rho_c$ reaches a upper limit for very small $\alpha$. In
contrast, in the case of positive $\alpha$, we find that $\rho_C$
increases dramatically and there exists a maximal value of
$\rho_c$, which indicates that although the leaders with large
degree play a key role in the cooperation, very strong influence
of leaders will do harm to the persistence of cooperation and make
individuals to be selfish. One can also find that the larger the
value of $b$, the larger the value of $\alpha$ corresponding to
the maximal $\rho_c$. Moreover, an interesting phenomenon is
observed in Fig. 2, that is when $b$ is small, positive $\alpha$
leads to better cooperative behavior than the negative one;
However, for large $b$, the system performs better when choosing
negative $\alpha$. These results imply that if the income of
defectors is only little more than that of cooperators, the
leader's effect will considerably enhances the cooperation; While
if the selfish behavior is encouraged in the system (large $b$),
the influential individuals will leads to the imitation of selfish
behavior and reduce the cooperator density in a certain extent. On
the contrary, restriction of leader's influence (negative $\alpha$
decreases the selected probability of large degree individuals by
their neighbors) results in better cooperation.

\begin{figure}
\scalebox{0.80}[0.80]{\includegraphics{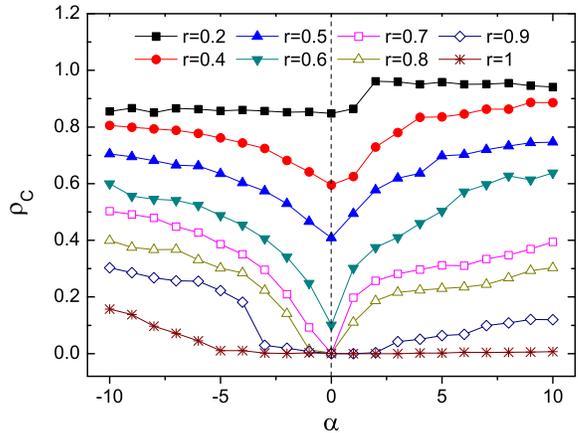}}
\caption{\label{fig:epsart} (color online). The cooperator density
$\rho_C$ of the SG as a function of parameter $\alpha$ for
different value of $r$.}
\end{figure}

In parallel, we investigate the effect of preferential learning
upon the SG. The simulation results are demonstrated in Fig. 3.
Similar to the PDG, $\rho_c$ is improved by the introduction of
preferential learning for nearly the entire range of $r$ from $0$
to $1$. In both sides of $\alpha=0$, $\rho_c$ reaches an upper
limit, which means that in the case of strong leader's influence
or without leaders, cooperation can be promoted to the highest
level for the wide middle range of $b$. Contrary to the PDG, for
very large $r$, the system still performs cooperative behavior,
which is attributed to the fact that the rule of SG favors the
cooperators, that is the cooperators yet gain payoff $1-r$ when
meeting defectors. Combining the above simulation results of both
the PDG and SG, we can conclude that the preferential learning
mechanism indeed plays an important role in the emergence of
cooperation.

\begin{figure}
\scalebox{0.80}[0.80]{\includegraphics{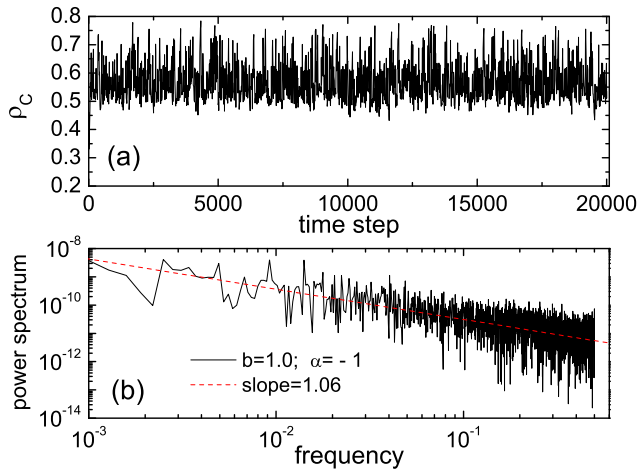}}
\caption{\label{fig:epsart} (a) Time series of cooperator density
$rho_C$ of the PDG for $b=1.0$ and $\alpha=-1$. (b) Power spectrum
analysis of (a).}
\end{figure}

In the following, we analyze the time series of the cooperator
density to give detailed description of the systems' evolutionary
behavior. We first study the PDG for negative value of parameter
$\alpha$. Surprisingly, for some specific values of $b$ and
$\alpha$, $1/f$-like noise is found. A prototypical example is
exhibited in Fig. 4. The $1/f$-like noise is observed frequently
in real-world systems, including healthy physiologic systems
\cite{heathy1,heathy2,heathy3}, economical systems
\cite{voss,bak}, as well as traffic systems \cite{Tadic}. However,
as far as we know, $1/f$ pattern hasn't been reported in the study
of evolutionary games. The $1/f$ noise denotes that the power
spectrum of time series varies as a power-law $S(f)\sim f^{-\phi}$
with the slope $\phi=1$. The spectrum exponent $\phi$
characterizes the nature of persistence or the correlation of the
time series. $\phi=2$ indicates zero correlation associated with
Brownian motion, where as $\phi=0$ corresponds to a completely
uncorrelated white noise. $\phi>2$ indicates positive correlation
and persistence i.e., if the process was moving upward (downward)
at time $t$, it will tend to continue to move upward (downward) at
future times $t'$; $\phi<2$ represents negative correlation and
anti-persistence. The intermediate case, $S(f)\sim f^{-\phi}$, is
a ``compromise" between the small-time-scale smoothness and
large-time-scale roughness of Brownian noise. Figure 4 (a) shows
the time evolution for $b=1.0$ and $\alpha=-1$, i.e. the case of
restriction of leader's influence. In this case, the density of
cooperators remains stable with frequently fluctuations around the
average value. Figure 4 (b) is the power spectrum analysis of the
time series of cooperator density. A prototypical $1/f$-like noise
is found with the fitted slope $\phi=1.06$. This result indicates
when the small degree individuals have large probability to be
followed, i.e., suppress the influential leader's effect, the
nontrivial long range correlation of evolutionary cooperative
behavior emerges. The similar phenomenon is also observed in the
SG for the case of negative $\alpha$, as shown in Fig. 5. The
emergence of the $1/f$ scaling is associated with the parameter
values $\alpha=-1$ and $r=0.5$. The discovered $1/f$ noise for
both two games is partly ascribed to the lack of influence of
leaders. Suppose that if the individuals with large connectivity
are chosen with large probability, their strategy will be easily
followed by a large number of persons, because those leaders
usually gain very high income. Since the influential ones only
take the minority, the evolutionary cooperative behavior will
mainly determined by the minority. Besides, the strategies of
those leaders are usually fixed due to their very high score, the
long range correlation of the fluctuation of cooperator density is
broken.

\begin{figure}
\scalebox{0.80}[0.80]{\includegraphics{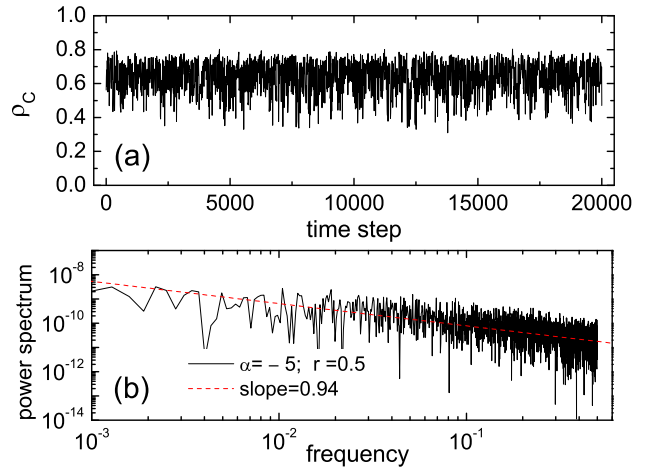}}
\caption{\label{fig:epsart} (a) Time series of cooperator density
$rho_C$ of the SG for $r=0.5$ and $\alpha=-5$. (b) Power spectrum
analysis of (a).}
\end{figure}

Then we investigate the evolutionary behavior of both the SG and
PDG in the case of positive $\alpha$. For the SG, when the
parameter $\alpha$ is close to zero, for arbitrary $b$, the level
of cooperation remains stable with relatively small fluctuations
around the average value. This property is remarkably changed for
large value of $\alpha$, which means the influence of leaders
becomes strong. As shown in Fig. 6, for $\alpha=5$ and $r=0.5$,
the equilibrium is punctuated by sudden drops of cooperator
density. After a sudden drop, the cooperation level $\rho_C$ will
gradually increase until $\rho_C$ reaches the average value. The
occurrence of these punctuated equilibrium-type behavior is
ascribed to the strong influence of a small amount of leaders. As
we have mentioned, the leader nodes usually get large payoffs,
thus they tend to hold their own strategies and are not easily
affected by their neighbors. However, those influential
individuals still have small probability to follow their
neighbors' strategies. If an event that a leader shift his
strategy to defector occasionally happens, the successful defector
strategy will rapidly spread from the leader to his vicinities.
Due to the connection heterogeneity of the scale-free networks,
i.e., the leaders have large amount of neighbors, the imitation of
a successful selfish behavior of the leader triggers the rapidly
decrease of cooperator density. After the occurrence of a sudden
drop, defectors become the majority and the selfish leader nearly
gain nothing. Then under the influence of the other leaders with
cooperate strategies, the cooperator density will slowly recover
to the steady state.

\begin{figure}
\scalebox{0.80}[0.80]{\includegraphics{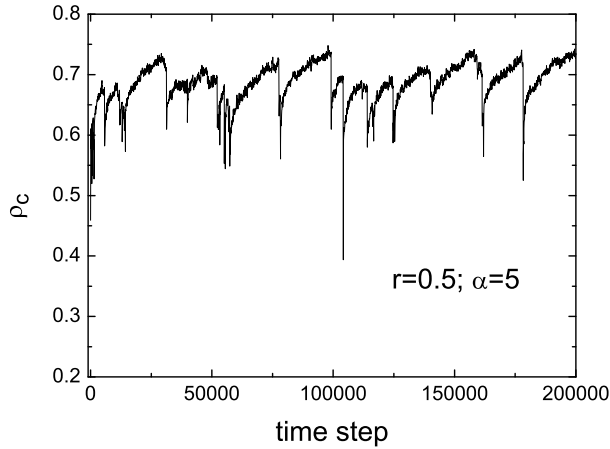}}
\caption{\label{fig:epsart} The time evolution of cooperator
density of the SG with $r=0.5$, $\alpha=5$ exhibits the punctuated
equilibrium-type behavior.}
\end{figure}

\begin{figure}
\scalebox{0.80}[0.80]{\includegraphics{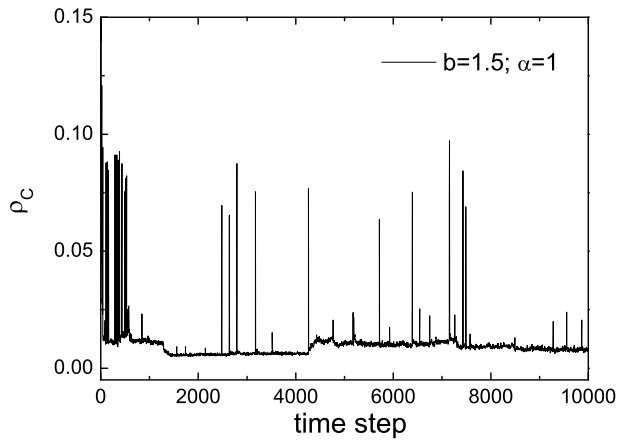}}
\caption{\label{fig:epsart} The sudden increase of cooperator
density of the PDG with $b=1.5$, $\alpha=1$.}
\end{figure}

The evolutionary behavior of the PDG for the positive $\alpha$
also exhibits nontrivial feature as shown in Fig. 7. Contrast to
the SG, the cooperation level shows some sudden increases. The
mechanism that induces the temporary instability of cooperator
density is the same as that of the sudden drops of the SG. The
strategy shift of influential nodes plays the main role in the
occurrence of the sudden increase. Opposite to the SG, the payoff
matrix of the PDG favors the defect behavior, thus the cooperation
level is quite low. An occasional strategy shift from defect to
cooperate of a leader will lead to the imitation of its neighbors
and a sudden increase occurs. However, the high cooperator density
is instable in the PDG for large $b$, hence the sudden increase
will rapidly decrease to the average value.

\section{conclusion and discussion}
We have investigated the cooperative behavior of the evolutionary
games resulting from the preferential mechanism. Comparing with
the cases of random selection, i.e., $\alpha=0$, preferentially
selecting large degree nodes or small degree ones can promote the
cooperator density of both the PDG and the SG over a wide range of
payoffs. For the cases of negative value of $\alpha$, the systems
perform the behavior of long range correlation, which is
quantified by the $1/f$ scaling of power spectrum. Interestingly,
in the case of positive value of $\alpha$, i.e., the large degree
nodes have high probability to be selected for imitation, the SG
exhibits a punctuated equilibrium-type behavior which is qualified
by the occasional occurrence of sudden drops. In contrast to the
SG, the PDG shows temporary instable behavior with the existence
of sudden increase. The mechanism that leads to the instabilities
of cooperation for both games are the strategy shift of
influential nodes and the imitation of their neighbors. The
instable behavior indicates that the strong influence of leader
individuals will do harm to the evolutionary cooperative behavior
of the systems. The present work implies that the existence of the
preferential learning mechanism plays an important role in the
emergence of cooperation in the heterogeneous networked systems.


\begin{thebibliography}{ref1}
\bibitem{cooperation} A. M. Colman, {\sl Game Theory and its Applications in the Social and Biological
Sciences} (Butterworth-Heinemann, Oxford, 1995).

\bibitem{gene} R. Dawkins, {\sl The Selfish Gene} (Oxford University Press, Oxford, 1989).

\bibitem{von} J. von Neumann and O. Morgenstern,  {\sl Theory of Games and Economic
Behaviour} (Princeton University Press, Princeton, 1944).

\bibitem{nash} J. Nash, Econometrica \textbf{18}, 155 (1950).

\bibitem{smith} J. Maynard Smith and G. Price, Nature (London) \textbf{246}, 15 (1973).

\bibitem{axelrod} R. Axelrod, {\sl The Evolution of Cooperation} (Basic books, New York,
1984).

\bibitem{sigmund} J. Hofbauer and K. Sigmund,  {\sl Evolutionary Games and Population
Dynamics} (Cambridge University Press, Cambridge, U.K.,
 1998).

\bibitem{http} C. Hauert, ``Virtuallabs: Interactive tutorials on evolutionary game
 theory", \href{http://www.univie.ac.at/virtuallabs}{http://www.univie.ac.at/virtuallabs}.

\bibitem{PD} R. Axelrod and W. D. Hamilton, Science \textbf{211}, 1390 (1981).

\bibitem{SG} R. Sugden, {\sl The Economics of Rights, Co-operation and
Welfare} (Blackwell, Oxford, U.K., 1986).

\bibitem{bio1} P. E. Turner and L. Chao, Nature (London)
\textbf{398}, 441 (1999).

\bibitem{bio2} P. E. Turner and L. Chao, Am. Nat. \textbf{161}, 497 (2003).

\bibitem{Nowak1} M. Nowak and R. M. May, Nature (London) \textbf{359}, 826
(1992); Int. J. Bifurcation Chaos Appl. Sci. Eng. \textbf{3}, 35
(1993).

\bibitem{Nowak2} M. Nowak and K. Sigmund, Nature (London)
\textbf{355}, 250 (1992).

\bibitem{Nowak3} M. Nowak and K. Sigmund, Nature (London)
\textbf{364}, 56 (1993).

\bibitem{Kim} B. J. Kim, A. Trusina, P. Holme, P. Minnhagen, J. S.
Chung, and M. Y. Choi, Phys. Rev. E \textbf{66}, 021907 (2002).

\bibitem{Szabo1} G. Szab\'{o} and C. T\"{o}ke, Phys. Rev. E
\textbf{58}, 69 (1998).

\bibitem{Szabo2} G. Szab\'{o} and C. Hauert, Phys. Rev. Lett.
\textbf{89}, 118101 (2002).

\bibitem{Szabo3} G. Szab\'{o} and C. Hauert, Phys. Rev. E
\textbf{66}, 062903 (2002).

\bibitem{Szabo4} G. Szab\'{o} and J. Vukov, Phys. Rev. E
\textbf{69}, 036107 (2004).

\bibitem{Szabo5} J. Vukov and G. Szab\'{o}, Phys. Rev. E
\textbf{71}, 036133 (2005).

\bibitem{Szabo6} C. Hauert and G. Szab\'{o}, Am. J. Phys. \textbf{73}, 405 (2005).

\bibitem{Hauert} C. Hauert and M. Doebeli, Nature \textbf{428}, 643
(2004).

\bibitem{Doebeli} M. Doebeli, C. Hauert and T. Killingback, Science \textbf{306},
859 (2004).

\bibitem{prl} F. C. Santos and J. M. Pacheco, Phys. Rev. Lett.
\textbf{95}, 098104 (2005).

\bibitem{SW} D. J. Watts and S. H. Strogatz, Nature (London) \textbf{393}, 440
(1998).
\bibitem{BA} A.-L. Barab\'asi and R. Albert, Science
\textbf{286}, 509 (1999).

\bibitem{heathy1} C.-K. Peng, S. Havlin, H.E. Stanley, and A.L.
Goldberger, Chaos \textbf{5}, 82 (1995).

\bibitem{heathy2} P.C. Ivanov, L.A.N. Amaral, A.L. Goldberger, S.
Havlin, M.G. Rosenblum, Z.R. Struzik, and H.E. Stanley, Nature
(London) \textbf{399}, 461 (1999).

\bibitem{heathy3} L.A.N. Amaral, P.C. Ivanov, N. Aoyagi, I.
Hidaka, S. Tomono, A.L. Goldberger, H.E. Stanley, and Y. Yamamoto,
Phys. Rev. Lett. \textbf{86}, 6026 (2001).

\bibitem{voss} R. F. Voss, {\sl 1/f noise and fractals in Economic time series} (Springer-Verlag, 1992).

\bibitem{bak} P. Bak, {\sl How Nature Works} (Oxford University Press, Oxford, 1997).

\bibitem{Tadic} B. Tadi\'c, S. Thurner, and G. J. Rodgers, Phys.
Rev. E \textbf{69}, 036102 (2004)

\end{thebibliography}
\end{document}